\documentclass[twocolumn, switch]{article} 

\usepackage{preprint}
\usepackage{amsmath, amsthm, amssymb, amsfonts}
\usepackage[numbers,square]{natbib}
\usepackage[utf8]{inputenc}	
\usepackage[T1]{fontenc}	
\usepackage{xcolor}		
\usepackage[colorlinks = true,
            linkcolor = purple,
            urlcolor  = blue,
            citecolor = cyan,
            anchorcolor = black]{hyperref}	
\usepackage{booktabs} 		
\usepackage{nicefrac}		
\usepackage{microtype}		
\usepackage{lineno}		
\usepackage{float}			

\usepackage{lipsum}		
\usepackage{url}

\usepackage[binary-units=true]{siunitx}
\usepackage[left]{eurosym}

\usepackage{newfloat}
\DeclareFloatingEnvironment[name={Supplementary Figure}]{suppfigure}
\usepackage{sidecap}
\sidecaptionvpos{figure}{c}

\usepackage{titlesec}
\titlespacing\section{0pt}{12pt plus 3pt minus 3pt}{1pt plus 1pt minus 1pt}
\titlespacing\subsection{0pt}{10pt plus 3pt minus 3pt}{1pt plus 1pt minus 1pt}
\titlespacing\subsubsection{0pt}{8pt plus 3pt minus 3pt}{1pt plus 1pt minus 1pt}

\usepackage{tikz,xcolor,hyperref}

\definecolor{lime}{HTML}{A6CE39}
\DeclareRobustCommand{\orcidicon}{
	\begin{tikzpicture}
	\draw[lime, fill=lime] (0,0)
	circle [radius=0.16]
	node[white] {{\fontfamily{qag}\selectfont \tiny ID}};
	\draw[white, fill=white] (-0.0625,0.095)
	circle [radius=0.007];
	\end{tikzpicture}
	\hspace{-2mm}
}
\foreach \x in {A, ..., Z}{\expandafter\xdef\csname orcid\x\endcsname{\noexpand\href{https://orcid.org/\csname orcidauthor\x\endcsname}
			{\noexpand\orcidicon}}
}

\title{HiRIS: an Airborne Sonar Sensor with a 1024 Channel Microphone Array for In-Air Acoustic Imaging}

\usepackage{authblk}

\author[1,2\thanks{\tt{dennis.laurijssen@uantwerpen.be}}]{Dennis Laurijssen\orcidA{}}
\author[1,2]{Walter Daems\orcidB{}}
\author[1,2]{Jan Steckel\orcidC{}}

\affil[1]{Cosys-Lab, Faculty of Applied Engineering, University of Antwerp, Antwerp, Belgium}
\affil[2]{Flanders Make Strategic Research Centre, Lommel, Belgium}

\begin{document}

\twocolumn[ 
  \begin{@twocolumnfalse} 
\maketitle


\begin{abstract}
Airborne 3D imaging using ultrasound is a promising sensing modality for robotic applications in harsh environments. Over the last decade, several high-performance systems have been proposed in the literature. Most of these sensors use a reduced aperture microphone array, leading to artifacts in the resulting acoustic images. This paper presents a novel in-air ultrasound sensor that incorporates 1024 microphones, in a 32-by-32 uniform rectangular array, in combination with a distributed embedded hardware design to perform the data acquisition. Using a broadband Minimum Variance Distortionless Response (MVDR) beamformer with Forward-Backward Spatial Smoothing (FB-SS), the sensor is able to create both 2D and 3D ultrasound images of the full-frontal hemisphere with high angular accuracy with up to 70\si{\decibel} main lobe to side lobe ratio. This paper describes both the hardware infrastructure needed to obtain such highly detailed acoustical images, as well as the signal processing chain needed to convert the raw acoustic data into said images. Utilizing this novel high-resolution ultrasound imaging sensor, we wish to investigate the limits of both passive and active airborne ultrasound sensing by utilizing this virtually artifact-free imaging modality. 

\end{abstract}
\vspace{0.35cm}

  \end{@twocolumnfalse} 
] 


\section{Introduction}
While microphone arrays have been around for more than 50 years~\cite{michel2006history}, the landscape of microphone array sensors and its technology have advanced tremendously in the last two decades with the rise of MEMS (micro-electro-mechanical system) technology. Furthermore, the last decade has given rise to many novel 3D in-air ultrasound sensors which allow the formation of acoustic images in 3D. These sensors hold great promise for robotic applications in harsh environments, as ultrasound signals are minimally affected by medium distortions such as dust, fog and water spray. However, the sensors developed in the past typically use a reduced aperture due to cost and complexity limitations, with microphone counts typically ranging from 1 to 64. These reduced apertures inevitably cause either artifacts in the resulting 3D images, or images with a limited dynamic range and spatial resolution.

Ultrasound signals often exhibit a large Helmholtz number in relationship to the environments where they are applied, implying that the reflected energy impinging on the sensor is mainly specular in nature. On the other hand, diffraction echoes should arise from acoustic theory~\cite{pierce2019acoustics}, but so far these echos have been mostly neglected due to their low intensity. In order to assess the relative importance of these echoes in real-world environments, as well as to investigate what the virtual upper limit is of ultrasound sensing in real-world environments, a sensor with a high spatial resolution and high dynamic range is necessary. 

This paper tries to address the need for a sensor with high spatial resolution and dynamic range, and introduces a dense, large aperture in-air ultrasound microphone array which should provide these high spatial resolutions, dynamic ranges and signal to noise ratios. The system is consists of 1024 synchronously sampled microphones, increasing the number of microphones of our previously developed systems by a factor of 32~\cite{kerstens2019, kerstens2017low, kerstens2019optimized}. This sharp rise in microphone channel count is achieved by leveraging a distributed hardware architecture, which is built upon a decade of ultrasound sensor development. In this paper, we demonstrate a successful implementation of this novel acoustic sensor, and demonstrate its functionality through both simulation and real-world measurements.

In order for the readers to accurately follow the developments, we encourage them to get familiar with our previous work in which we describe in detail the development of a single 32-channel microphone array ~\cite{kerstens2019, verellen2020}, as the sensor in this paper consists of a distributed version of that single 32-channel module. However, this paper still stands on its own, allowing the reader to follow the development of the data-acquisition methods and signal processing techniques and understand the performance analysis of the system where we compare it with our previously developed 32-channel microphone array~\cite{kerstens2019, verellen2020}. 

In the pages that follow, we will touch on the design choices that were made to achieve the hardware architecture of the developed ultrasound sensor unit that we named the \emph{High Resolution Imaging Sonar (HiRIS) sensor}, together with a more detailed description of the implementation. In the subsequent section, the data acquisition and signal processing are described followed by a section on the experimental setup and its results. In the final section, we will present the conclusions of the proposed system and its envisioned applications as future work.


\section{Hardware Architecture}
Achieving the envisioned objective of constructing a synchronized ultrasound sensor array featuring 1024 microphone channels, coupled with a versatile yet timely data transfer interface, poses a nontrivial challenge. It requires considering several trade-offs in different design aspects, such as the choice of components and their associated costs, design time influenced by familiarity with a platform, as well as the time allocated for implementation and testing. This section aims to delve into the deliberations behind these design choices, exploring considerations related to component types, cost implications, design familiarity impact on time, and the overall implementation process. Additionally, we will introduce the selected implementation and provide an overview of the proposed system.

The hardware design of the HiRIS sensor is a highly complex one, which warrants its own extensive description as the devil is in the details. Indeed, the overall system consists of 1024 microphones, 33 microcontrollers, 4963 electronical components, distributed over 2 PCBs, and over 127m of PCB traces. The road to a successful implementation of such a system is riddled with pitfalls, which we aim to clarify in the subsequent sections.

\begin{figure*}
    \centering
    \includegraphics[width=\linewidth]{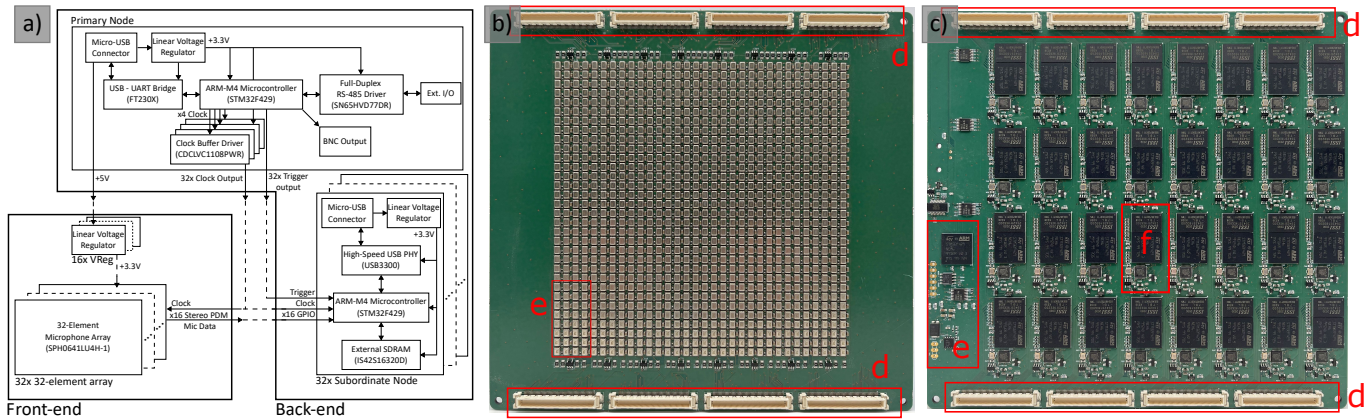}
    \caption{Overview of the HiRIS hardware architecture. Panel a) show a schematic representation of the system architecture, distinguishing between the front-end and back-end PCBs. On the front-end, there are 32 groups of 32 microphones (each arranged in an 8x4 grid). The back-end PCB has one primary node which does clock distribution, triggering and synchronization of the subordinate nodes. The subordinate nodes each sample a 32-channel microphone group, using 16 IO pins with the microphones operating in dual-channel stereo mode.  Panel b) shows the assembled front-end PCB measuring 180\si{\milli\meter} by 170\si{\milli\meter}. Box e) indicates a single 8x4 group of microphones. Boxes labeled d) indicate the interconnect connectors between the front-end and back-end PCBs. Panel c) shows the assembled back-end PCB, which has the same dimensions as the front-end PCB. Panel e) indicates the primary node, and box f) shows a single subordinate node.} 
    \label{fig:hiris}
\end{figure*}

\subsection{Design Choices}
Over the last decade, the embedded products market has witnessed a notable surge in diversity, propelled by swift technological advancements. The integration of highly capable and feature-rich (ARM) microcontrollers, along with Field-Programmable Gate Arrays (FPGAs) and System on Chips (SoCs), has been instrumental in enhancing the capabilities of embedded sensor systems. The emergence and widespread adoption of Single Board Computers (SBCs), coupled with the proliferation of Internet of Things (IoT) devices, have been spurred by the demand for technological progress in the era of Industry 4.0. Concurrently, the growing community of online hobbyists in the domain of embedded electronics hardware has contributed to the development of tools and libraries, facilitating the rapid creation of embedded platforms.
\newline\newline
While the three aforementioned types of embedded devices, being FPGAs~\cite{kerstens2017low, kerstens2019, allevato2020real, allevato2022air, allevato2023two}, SoCs~\cite{verreycken2017distributed, verreycken2018passive, verreycken2021bio} and ARM microcontrollers~\cite{laurijssen2019flexible, verellen2020urtis, kerstens2019optimized, laurijssen2023bio, laurijssen2016flexible}, have been used for the construction of high-resolution ultrasound sensing arrays, each of these device types have their distinct advantages and disadvantages. 
\newline\newline
Field-Programmable Gate Arrays (FPGAs) offer notable advantages in terms of flexibility and customization. Their field-configurable nature, in combination with very high GPIO pin counts, allows for their rapid adaptation to diverse tasks, especially in real-time applications with very tight timing constraints. However, the complexity of FPGA design, coupled with the relatively higher power consumption and component cost can be considered drawbacks. Indeed, developing complex hardware designs in FPGAs is complicated due to the need for tight timing closures, in order to yield stable data-acquisition systems.

Systems-on-a-Chip (SoCs) integrate multiple components, processing cores and peripherals onto a single chip, streamlining the design and reducing the need for external components. This leads to space and power efficiency. While the large amount of diverse peripherals on the SoC is attractive, the amount of customization options are significantly more limited when compared to FPGAs. Therefore, complexity in the envisioned design may lead to significant challenges during the development. On the other hand, timing closure is guaranteed by design, leading to far less potential for race conditions compared to FPGA-based designs.

ARM microcontrollers excel in power efficiency and simplicity, due to their standardized architecture and interface design. The low cost of ownership, coupled to wide industry adoption make them accessible for a wide range of applications, from automotive, over consumer goods, and indeed, to high-speed data-acquisition. This high degree of standardisation leads to a reduced customization potential when compared to FPGAs, or the integrated capabilities of SoCs, limiting their suitability for certain high-performance or specialized tasks.

Despite the apparent drawbacks of ARM-based microcontroller systems, we deemed this to be the most promising candidate for the development of the hardware architecture for HiRIS. While the other two options (being SoCs and FPGAs) are certainly viable options for implementing such a systems, we chose for an ARM-based architecture, because of the fact that a) the peripherals on the chosen ARM platform are ideally suited for our intended application, b) a distributed architecture is more error robust than a single monolithic implementation, and c) our group is well versed in the development of ARM-based systems, which is a non-neglectable reason for choosing a particular approach.

\subsection{Distributed Architecture}
When considering the design of complex hardware systems, it often pays off to approach the implementation using a distributed architecture. Indeed, when using a distributed architecture, robustness increases due to the lack of single point-of-failures. In the case of the embedded hardware design of HiRIS, it can be beneficial to split up the hardware over multiple printed circuit boards (PCBs) that are tied together using one or multiple appropriate connectors. The hardware components can be grouped by functionality and can hence be isolated in the design process, which in turn can have advantages during the implementation and testing phase. This is especially important for testing individual boards with high component counts, as this distributed approach allows them to be tested without inducing dangerous voltages or currents to other parts of the device. Furthermore, sections of the design can easily be redesigned if deemed necessary after testing (i.e., it facilitates an iterative design approach), without having to reassemble the non-faulty parts of the system. 
\newline
This modular approach also has the added benefit of being able to make use of an extra spatial dimension in the hardware design by connecting multiple PCBs on top of each other, reducing the surface of the total design to its volume. When designing acoustic array sensors of the proposed complexity encountered in HiRIS, we often separate the microphones and some of their essential peripherals to a front-end PCB, and place the rest of the electronic components to a so called back-end PCB. As a beneficial side effect, this creates the potential for leaving front-facing side of the front-end PCB component-less, which is essential for eliminating distorting multipath effects in the acoustic reception pathways.
\newline
\newline
Another advantage of the distributed architecture can be found in reusing known, verified and tested schematic and component layouts, used extensively in previous designs (i.e. the so-called battle-tested designs). By reusing parts of both the schematics and component layouts from previously built hardware, these parts can be distilled into design blocks, which then can be combined in the larger overarching design. These design blocks allowed us to quickly create a distributed hardware architecture of 32 microphone nodes by 33 microcontroller nodes, where every 32-element microphone node on the front end PCB is connected with an ARM microcontroller node with its peripherals on the back-end PCB. To orchestrate the 32 microcontroller nodes, an additional primary node was added on the back end. Employing this distributed design method of reusing existing design blocks has proved to be a highly productive and cost-efficient design methodology.

\subsection{Front end}
The front-end board mainly incorporates 1024 Knowles SPH0641LU4H-1 MEMS microphones, sixteen AP2112K-3.3V linear voltage regulators that convert +5\si{\volt} to +3.3\si{\volt} that is used as the power source for the sensing elements. While each of the microphones typically only consumes 850\si{\micro\ampere}, we have provided for ample headroom in the power budget of the voltage regulators, which is why these regulators each power a group of 64 microphones.
\newline\newline
The aforementioned MEMS microphones are configured in a 32-by-32 uniform rectangular array with regular grid spacing of 3.9\si{\milli\meter}. When designing phased arrays, the impact of its configuration on the resulting frequency dependent steerable beam directivity pattern is determined by the sensing element spacing $\mathit{d}$. We can apply the Nyquist theorem to the spatial domain~\cite{dmochowski2008spatial} and determine that grating lobes~\cite{konetzke2015phased} are introduced into the directivity pattern, as spatial aliases, when the sensing element spacing $d \geqslant \frac{\lambda}{2}$. Rearranging this simple equation, we can calculate the maximum frequency $f_{\mathit{max}}$ for which this array geometry can be used for beam steering without spatial aliases: 

\begin{align*} 
d = \frac{\lambda}{2} = \frac{v}{2f}  
\quad\Leftrightarrow\quad
f = \frac{v}{2 d}
\end{align*}

assuming the speed of sound in air $\mathit{v}$ is approximately 343\unit[per-mode = symbol]{\meter\per\second}, $f_{\mathit{max}}$ is found to be 43.974\si{\kilo\hertz}.
\newline\newline
Besides the small form factor, low power consumption and a frequency response curve reaching far into the ultrasonic spectrum~\cite{kerstens2017low}, the key advantage of the SPH0641LU4H-1 microphones is their built-in \si{\Sigma\Delta} ADC (analog-to-digital converter) that converts the captured acoustic wavefront into a PDM (pulse density modulation) 1-bit signal. The aforementioned \si{\Sigma\Delta} ADC uses an external clock signal of 4.5\si{\mega\hertz} for sampling the analog signals to their digital 1-bit representation. Given the relatively large wavelength of an electrical wave of 4.5\si{\mega\hertz} in copper being approximately 44.4\si{\meter}, we can distribute these clock signals in phase to all 1024 microphones, thus allowing simultaneous sampling all of the microphones distributed over the PCB. This synchronous sampling is important for the subsequent operation of the sensor in an array fashion, which often leads to significant complications in RADAR-based sensing applications~\cite{liang2019high, yang2013time, chaudhary2015characterization}. 
\newline\newline
Using microphones with a built-in 1-bit ADC has as a major advantage a significant reduction in board complexity. Indeed, if microphones with an analog voltage output were to be used, each of these microphone signals would need amplification and a dedicated ADC chip, which adds a significant amount of complexity (as demonstrated in our earlier designs~\cite{steckel2012broadband, steckel2013batslam}). Interfacing with 1-bit signals can be easily done using a wide GPIO register on a microcontroller. Further reduction of the necessary GPIO lines can be realized by utilizing the stereo-capability of the used PDM microphones. When using this implemented feature, one microphone will deliver its data on the rising edge of the clock signal where the other will deliver it on the falling edge. The latching of the data will occur on the opposite edges of the clock signal of the microphones. This stereo setup halves the number of data lines that are required from 1024 (without using the stereo feature) to 512 (when using the stereo feature). It should be noted that the while the induced phase differences of sampling on both the rising and falling edge of the data sampling is 180$^\circ$, this equates to 10\si{\micro\second}, which equates to a negligible phase difference when compared to the targeted acoustic signals between 25\si{\kilo\hertz} and 100\si{\kilo\hertz}.
\newline\newline
In addition to the aforementioned voltage regulators, microphones and various passive components e.g. resistors and decoupling capacitors, eight high density FX10A-120P connectors can also be found on this PCB to connect the power, multiple synchronous clock lines and data lines to the back-end PCB of this design. These connectors ensure a high-fidelity link of the digital signals from the front end to the back end, ensuring robust operation of the sensor during field-trials.

\subsection{Back end}
The back end of the HiRIS sensor can be split up in a primary node and 32 subordinate nodes. The latter are 32 identical design blocks with the STM32F429 ARM Cortex M4 microcontroller at its core, in combination with an IS42S16320D external SDRAM memory of 64\si{\mega\byte}, a USB3300 high-speed USB PHY, an AP2112K-3.3V linear voltage regulator, a micro USB connector and passive connectors for decoupling and impedance matching of the transmission lines. These nodes each have 16 GPIO pins respectively connected to 32 microphones in their stereo configuration, together with a separate GPIO pin acting as clock signal. Based on the rising and falling edges of the clock signal, the 16 GPIO pins are synchronously sampled and temporarily stored in memory until the next clock period. Upon receiving a trigger signal from the primary node, the subordinate nodes will store these temporarily into the SDRAM until a predetermined number of samples have been recorded. Given the 64\si{\mega\byte} of SDRAM memory capacity that every subordinate node has and a data stream of 18\unit[per-mode = symbol]{\mega\byte\per\second} per 32 microphones that are connected to it, a single continuous measurement of approximately 3.55\si{\second} can be recorded.
\newline
\newline
The primary node, that also has a STM32F429 ARM Cortex M4 microcontroller at its core, uses a single timer peripheral that generates 4 synchronous square wave signal outputs at 4.5\si{\mega\hertz} that in turn each are connected to a Texas Instruments CDCLVC1108PWR low jitter, 1:8 LVCMOS fan-out clock buffer IC, effectively yielding 32 synchronous 4.5\si{\mega\hertz} clock signals. These clock signals are distributed to the 32 subordinate node on the back end and their respective microphones on the front-end PCB. This ensures on-the-clock-true sampling of all microphones, which is important for the subsequent processing pipeline. While every subordinate node is equipped with a bi-directional high speed USB interface for establishing a connection with a computer, this interface does not suffice to initiate a synchronous start of a measurement for all nodes, due to the unpredictability of the timing of data transfer over 32 separate USB channels. In order to establish a synchronous trigger to start a single measurement, the primary node uses 16 GPIO pins that are connected to the 32 subordinate nodes to initiate measurements on the latter by synchronously asserting a short pulse on these pins. Since the primary node does need a communication interface to a computer, it does not need to transfer large quantities of data in a short amount of time. Therefore an FTDI FT231X USB-to-UART bridge has been used in the primary node to establish a low-speed but reliable interface. 
\newline\newline
Besides a USB connection to the primary node for initiating measurements, an external TTL-input can be used to trigger measurements, which allows for easy integration of the HiRIS sensor in measurement pipelines. For increased robustness, e.g. long cable lengths or noisy environments, the option of using differential signaling for external I/O was chosen by incorporating a SN65HVD77DR RS-485 driver into the design. Since this is a full-duplex interface, a pulse can also be generated to trigger external devices along with the subordinate nodes. 
\newline\newline
While the HiRIS is designed as a passive measurement device, a BNC connector was also fitted to the back end that is connected to the analog DAC-output of the primary node. The DAC peripheral can be triggered simultaneously with the subordinate nodes where it will generate an analog signal on its output based on a sequence that was either pre-defined in the firmware or uploaded to the primary node through its USB interface. This enables us to use this sensor as a high-channel pulse-echo sonar device when combined with an external amplifier and ultrasound transducer, similar to the sensors described in our earlier work~\cite{kerstens2019, kerstens2017low, kerstens2019optimized, verellen2020urtis, steckel2012broadband, steckel2013batslam}.

\subsection{Physical Realisation of the HiRIS Sensor}
The proposed 1024-microphone ultrasound array sensor, referred to as the HiRIS sensor (High-Resolution Imaging Sonar), comprising of the front and back end PCBs measures 180\si{\milli\meter} by 170\si{\milli\meter} by 20\si{\milli\meter} (including its protruding micro USB connectors). The two boards combined feature nearly 5000 components, 127\si{\meter} of PCB traces and costs \EUR{3962} to produce. Once powered up and operational, the overall sensor system consumes approximately 65\si{\watt}. This provided an unanticipated heat output of the sensor, which called for an adequate cooling solution. Indeed, when measuring the surface temperature of the PCBs, we noticed areas that reached up to 80\si{\degreeCelsius} in a room with an ambient temperature of approximately 22\si{\degreeCelsius}. While electronics are often rated to cope with higher temperatures it was deemed that including a method for passively cooling the sensor would be beneficial for its lifespan and the safety of its users. As a cooling solution, a copper slab of 250\si{\milli\meter} by 170\si{\milli\meter} by 5\si{\milli\meter} was used between the front end and back end with a thermal interface on both sides for providing an optimal surface contact between the components and copper. To further increase the heat dissipation of the cooling solution, extra heat sinks were bolted onto the protruding ends of the copper slab. The effect of this cooling solution yielded a temperature decrease of 35\si{\degreeCelsius} with maximum surface temperatures reaching up to 45\si{\degreeCelsius}.
\newline\newline
As mentioned in the previous subsection, the HiRIS sensor can be expanded by connecting external devices through its external I/O or BNC connections but could also be further expanded with an additional PCB that stacks on the backside of the back end. This envisioned additional PCB would incorporate multiple USB3 hub ICs in order to reduce the amount of cable clutter. Another feature that would be integrated is a JTAG SWD programmer in combination with a multiplexer to alleviate the tedious work of plugging and unplugging the programmer when pushing firmware changes to the microcontrollers on the back end.


\section{Data acquisition and Processing Chain}
In this section, we will detail the process of initiating and capturing a set of waveform data from the microphone array, and the subsequent processing using a bank of adaptive spatial filters (MVDR beamforming) for 3D image generation.

\begin{figure}
    \centering
    \includegraphics[width=0.95\linewidth]{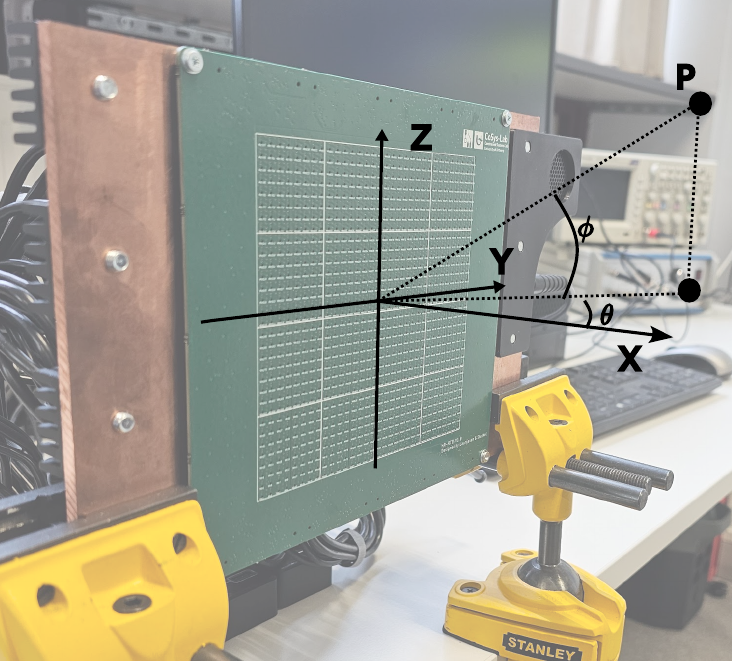}
    \caption{Coordinate system in relationship to the HiRIS sensor, showing the X, Y and Z axis, and the azimuth angle $\theta$ and elevation angle $\phi$.}
    \label{fig:coordinatesystem}
\end{figure}

\subsection{DAQ}
The HiRIS sensor comprises of 33 microcontrollers (1 primary node and 32 subordinate nodes), each connected over a USB2.0 connection to a host PC. To aggregate all the USB connections, a chain of USB3 hubs is used, which aggregates all the USB connections to a single USB3.0 connection. The USB protocol is a CDC Virtual Com Port (VCP) emulation~\cite{USBCDC}, each initializing a virtual serial port on the host PC. A custom Python script using the Multiprocessing API~\cite{singh2013parallel}, looks for specific serial ports connected to the system with specific Vendor ID and Product ID combinations and opens all these ports, which allows for bidirectional communication with all the sensor nodes.
\newline\newline
As stated before, the primary-subordinate architecture of our sensor implies that the single primary node of the HiRIS sensor listens to a command originating from a controlling PC over the VCP. In turn, the primary node asserts a trigger pulse to the 32 subordinate nodes, which each perform a measurement of a set duration (typically 70\si{\milli\second}). This data is then sent by each subordinate node over its serial port to the host PC, which combines all the data and stores them in a binary format on a mass storage device for subsequent processing. A single measurement of 70\si{\milli\second} is around 1.25 \si{\mega\byte} of data per subordinate node, equating to 40\si{\mega\byte} for a single measurement. Reworking this implies a datastream of 1\si{\giga\byte} per second when measuring with a 100 percent duty cycle. In practice, a duty cycle of 20 percent is more realistic, leading to a datarate of 200\si{\mega\byte} per second.

\subsection{Image Formation using MVDR}
\newcommand\PDM{\mathit{PDM}}
\newcommand\MF{\mathit{MF}}
\newcommand\MVDR{\mathit{MVDR}}
To process the massive amount of microphone data into a spacial spectrum, we follow an approach similar to the one outlined in~\cite{verellen2020urtis}. The microphone signals are PDM modulated using single-bit \si{\Sigma\Delta} modulation. In order to demodulate these signals, we pass them through a low-pass filter, and decimate the resulting signal:
$$
s_{M,i}(t) = h_{\PDM} * s_{\PDM,i}(t)
$$
The cutoff frequency of $h_{\PDM}$ equals to 100\si{\kilo\hertz}, which is well before the rise of the colored quantization noise, induced by a noise shaping on the MEMS microphones. This results in the $i$-th microphone signal $s_{M,i}$, of which there are 1024 in the case of the HiRIS sensor. These 1024 microphone signals are all passed through a matched filter. The base signal $s_b(t)$ is the emitted signal in the case of an active sonar measurement (where the sensor emits a signal), or a Dirac delta function in case of a passive measurement (where the sensor listens to environmental signals:
$$
s_{\MF,i}(t) = \mathcal{F}^-1 \bigg[ \mathcal{F}[s_b(t)]^* \cdot \mathcal{F}[s_{\MF,i}(t)]\bigg ]
$$
Next, these signals are converted into a time-frequency distribution using the short-time Fourier Transform (STFT), yielding a spectrum $S_{\MF,i}(t,f)$ for each $i$-th microphone signal. We choose a certain operating frequency $f$, in this case 42 kHz, and select the column of the STFT according to that frequency. This yields a complex signal $x_i(t)$ for each microphone. These signals are all concatenated into an observation matrix called $X(f,t)$:
$$
X(f,t) = \begin{bmatrix} x_1(f,t) & x_2(f,t) & ... & x_k(f,t) \end{bmatrix}
$$
For the spatial filtering, we apply a function on $X(f,t)$, depending on the kind of beamforming we want to achieve. We limit the scope of this paper to Minimum Variance Distortionless Response beamforming (MVDR)~\cite{verellen2020,verellen2020urtis, van2002optimum}. We use forward-backward spatial smoothing with diagonal loading in order to overcome the limitations posed by the sonar sensing modality in that only a single snapshot can be used to perform spatial filtering. Indeed, in many applications such as radar or mobile communications, multiple snapshots are available. However, due to the limited speed of sound, this is not possible in the HiRIS application~\cite{verellen2020,verellen2020urtis}. Therefore, we apply spatial smoothing by selecting sub-arrays of size 28x28, yielding 25 virtual snapshots from the 25 subarrays formed during the spatial smoothing process. From this, we build the sample covariance matrix $R_{b}$, and then calculate the weights of the MVDR beamformer:
$$
w_{\MVDR}(\psi)=\frac{ R_b^{-1} \cdot A(\psi)}{A(\psi)^H \cdot R_b^{-1} \cdot A(\psi)} 
$$
where $A(\psi)$ is the array manifold matrix for spatial direction $\psi$ given the subarray geometry and frequency of operation. Using these MVDR-weights $w_{\MVDR}$ we can then apply the spatial filter on the observation matrix $X(f,t)$ by complex multiplication, yielding a beamformed signal $x_\psi(t,f)$ in direction $\psi$. To obtain spatial images of the surrounding of the sensor, various sampling strategies for $\psi$ can be derived ~\cite{reijniers2020optimized}. Indeed, in order to form a 2D image in the horizontal plane, we sample the azimuth angle $\theta$ in a regular manner from -90$^\circ$ to 90$^\circ$ (e.g., in steps of 1$^\circ$), while keeping the elevation angle $\phi$ constant and zero. For 3D images, we sample the direction vector $\psi$ uniformly on a sphere using a recursive zonal sphere partitioning algorithm~\cite{leopardi2006partition}, as this is the optimal sampling strategy for 3D scenes without prior knowledge.


\section{Verification of HiRIS}
\subsection{Simulation of Point Spread Functions}
\begin{figure}
    \centering
    \includegraphics[width=0.7\linewidth]{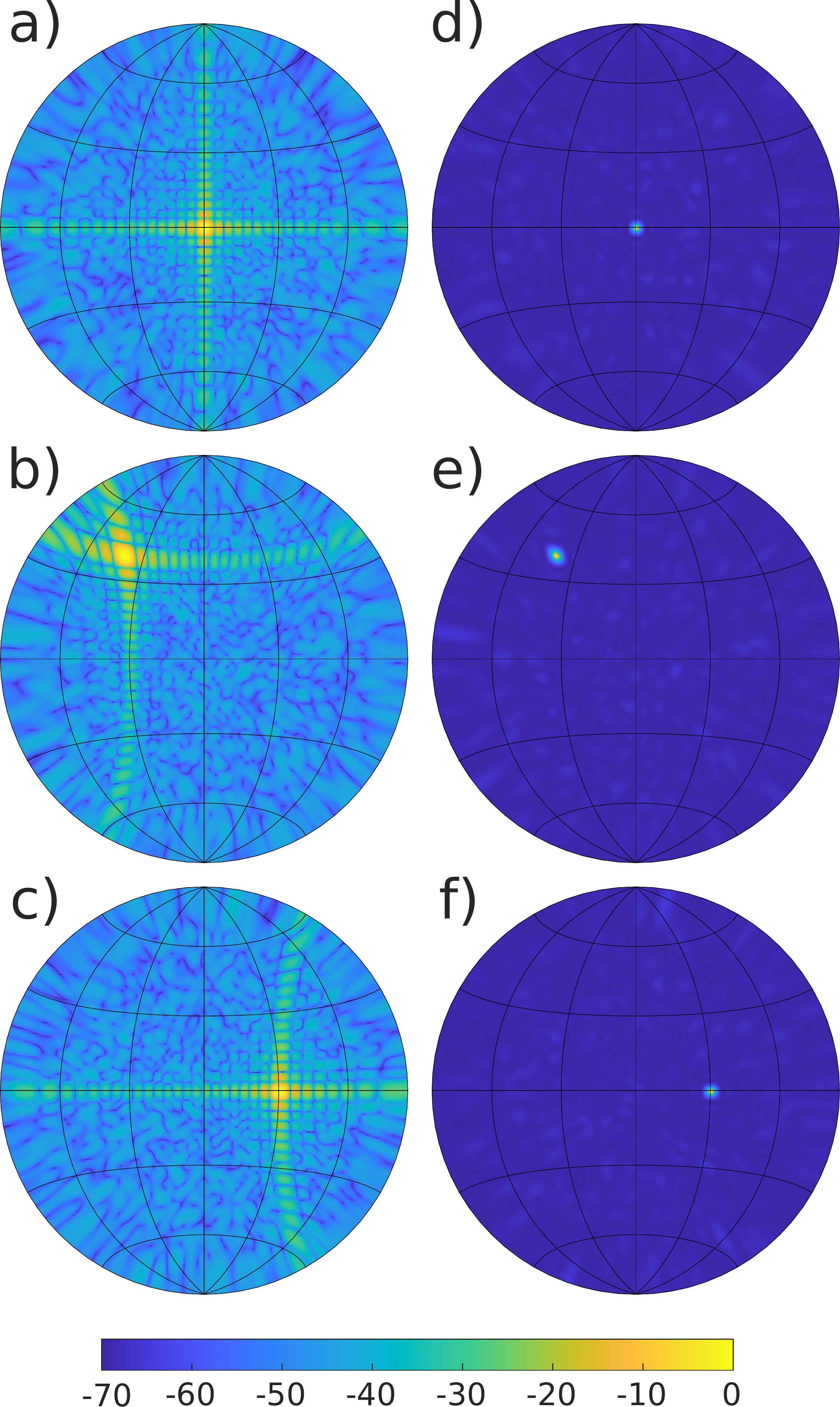}
    \caption{Point Spread Functions of point sources placed at different spatial locations: panel a \& d) $(\theta,\phi)=(0^{\circ},0^{\circ})$, panel b \& e) $(\theta,\phi) = (30^{\circ},0^{\circ})$, and panel c \& f) $(\theta,\phi)=(-45^{\circ},45^{\circ})$. Panels a-c) show the response of the system using Bartlett beamforming, and panels d-f) show the response when using the MVDR beamformer. The PSFs are shown on a logarithmic scale.}
    \label{fig:psfComparison}
\end{figure}

In order to verify the operation of the HiRIS sensor, a simulation model of the sensor was built, following the equations derived in~\cite{van2002optimum}. We calculate a so-called Point-Spread Function~\cite{steckel2015sonar, kerstens2019optimized} of the sensor system, which describes the image obtained by the sensor in response to a Dirac-like point source in space. We placed the point source in three spatial locations, defined by their azimuth angle ($\theta$) and elevation angle ($\phi$): $(\theta,\phi) = (0^{\circ},0^{\circ})$, $(30^{\circ},0^{\circ})$ and $(-45^{\circ} ,45^{\circ})$. The resulting Point Spread Functions (PSFs) can be found in figure \ref{fig:psfComparison}. Panels a-c show the point-spread function calculated using conventional Bartlett beamforming, and panels d-f show the PSF when using MVDR beamforming with spatial smoothing, a sub-array size of 28x28 (yielding 25 subarray snapshots), a signal to noise ratio of 5\si{\decibel} and a diagonal loading of 0.1. What becomes clear from these point-spread functions is their extremely narrow opening angle, and, especially in the case of the MVDR beamformer, excellent peak to sidelobe ratio (approaching 70\si{\decibel}), which in turn will allow the construction of high-resolution acoustic images.

\subsection{Real-world validation: Setup}
\begin{figure*}
    \centering
    \includegraphics[width=\linewidth]{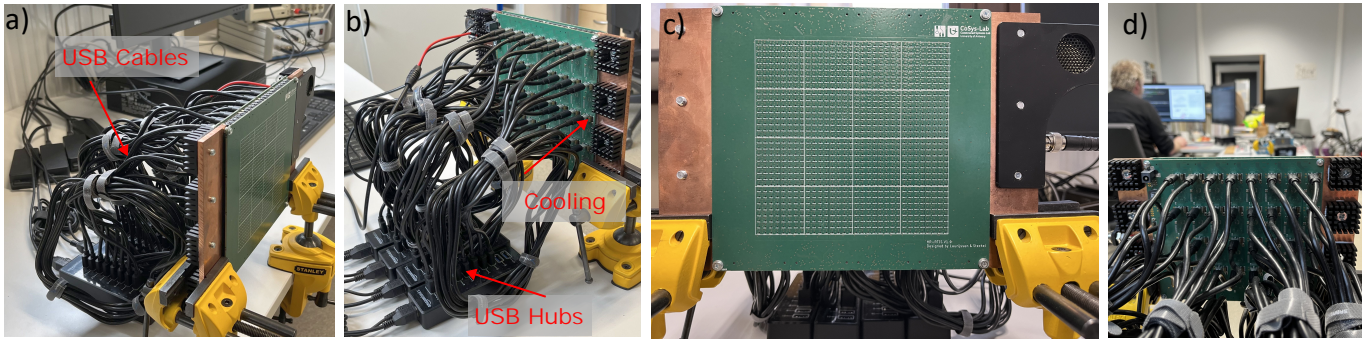}
    \caption{The realized prototype of HiRIS. Panel a) shows the front-view of the sensor with the microphone port-holes, and the 33 USB cables used to connect the nodes to the USB hubs. Panel b) shows the backside of the back-end PCB, with the USB cables connecting all the nodes to the USB hubs,and shows the copper cooling solution provisioned for heat management. The four USB hubs are then connected to an aggregate USB hub, which is connected to the host computer. Panel c) shows the front-view of the HiRIS sensor, where the component-less front-side is visible, with the exception of the holes for the bottom-mounted MEMS microphones. Panel d) shows the cluttered office space which has been ensonified during the active measurement experiment.} 
    \label{fig:exp_setup}
\end{figure*}

The realized prototype of HiRIS can be seen in figure \ref{fig:exp_setup}. Panel a) shows the front-view of the sensor with the microphone port-holes and the copper slabs used for cooling. Panel b) shows the backside of the back-end PCB, with the USB cables connecting all the nodes to the USB hubs. These four USB hubs are then connected to an aggregate USB hub, which is connected to the host computer.

\subsection{Real-world validation: Passive measurement}
In order to validate the Point-Spread Function of the realized prototype, we performed a passive acoustic measurement using an 40-\si{\kilo\hertz} ultrasonic source placed in front of the microphone array, emitting this pure tone at approximately 70\si{\decibel} SPL. A recording was made and the resulting data processed by the processing pipeline outline previously. The resulting images of the Point-Spread functions can be seen in figure \ref{fig:psfComparisons}. Panels a and b) show the resulting PSF when the data is processed using a broadband time-domain beamformer (delay-and-sum~\cite{kerstens2019}), both on a logarithmic (a) and linear (b) scale. Panels c) and d) show the response of the system when using an MVDR beamformer, outlined in the previous section. High Peak-to-sidelobe ratios can be noted in these PSFs, however with some deviations from the simulated PSFs shown in figure \ref{fig:psfComparisons}. The reason for this discrepancy is most-likely the slight phase differences between the simulation models (which assumes a zero-phase transfer function of each microphone) and the real-world microphones (where a slight variation might occur in the phase response). However, calibration techniques to compensate these transfer function differences exist and can be easily incorporated into the processing pipeline~\cite{chaudhary2015characterization,schuss2019large, he2021impact, agrawal2003calibration}. Finally, panels e) and f) show the response of the system to conventional Bartlett beamforming, again corresponding to the responses simulation in figure \ref{fig:psfComparison}. A much higher 'noise floor' can be observed, caused by the more prominent sidelobes present in conventional beamforming.

\subsection{Active Measurements}
\begin{figure*}
    \centering
    \includegraphics[width=\textwidth]{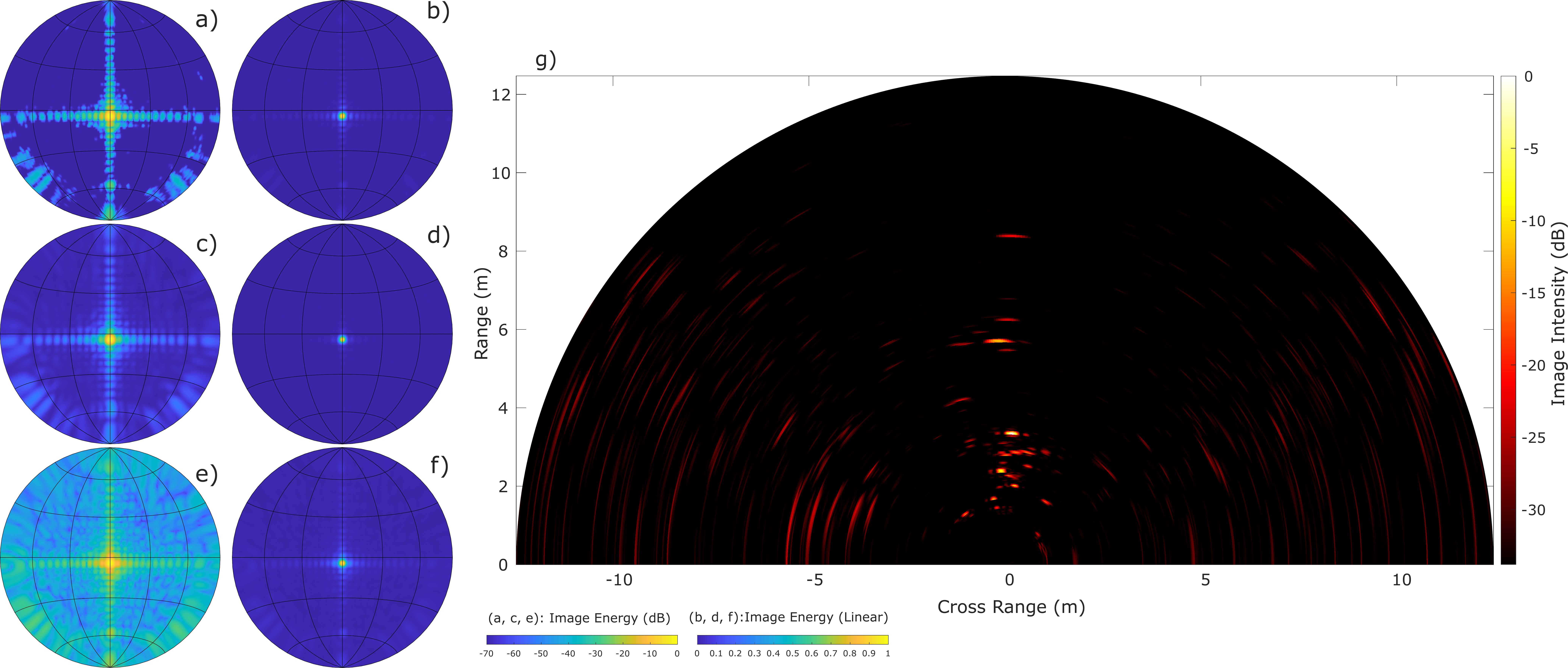}
    \caption{Experimental results of the HiRIS sensor. Panels a-c show the response of a 40-\si{\kilo\hertz} source placed in front of the HiRIS sensor, using various processing techniques (a: Delay and Sum, b: MVDR and c: Bartlett beamforming), on a logarithmic scale. Panels b-f) show the response on a linear scale. Panel g) shows the B-mode image of a scene ensonified using a broadband chirp, and processed using the algorithm described in this paper.} 
    \label{fig:psfComparisons}
\end{figure*}

As a final experiment, we performed an active measurement. In this case, the HiRIS sensor uses a Senscomp 7000 transducer~\cite{kerstens2019, steckel2015sonar, steckel2012broadband,steckel2013batslam,kerstens2019optimized,laurijssen2015three} to emit a broadband hyperbolic chirp. This chirp is generated by the DAC of the primary node and amplified using a custom high-voltage amplifier to a signal with an amplitude of 200\si{\volt}, superimposed on a bias of 200\si{\volt}. This emitted signal is reflected by the environment, and the subsequent reflections are recorded by the microphones. The signals are then processed into a 2D image using the method outlined before. The resulting 2D image can be found in figure \ref{fig:psfComparisons}, panel g). It shows the Cartesian representation of the 2D polar image (range and azimuth), on a logarithmic intensity scale. These images are often referred to as Energyscapes~\cite{steckel2012broadband} or B-mode images~\cite{sassaroli2019image, matrone2014delay}. They show narrow responses both in angle (due to the MVDR beamforming using the 1024 microphone array), as well as narrow range localization (due to the broadband signal used in the matched filtering step), and a high signal-to-noise ratio due to the high number of microphones used.

\section{Discussion and Future Work}
In this paper, we have presented HiRIS, the High Resolution Imaging Sonar, a sonar sensor with 1024 microphones. This microphone array is, to the best of the knowledge of the authors, the largest microphone array developed for ultrasound imaging in air to date of writing. We detailed the hardware architecture of the HiRIS sensor, indicating design choices and potential pitfalls when reproducing the hardware system. We provided a reasoning on why certain design choices have been made, and which can be used to inform future decisions when building similar hardware systems. Furthermore, we detailed the data-acquisition pipeline and signal processing approach, and tried to develop an intuition about the scales involved when dealing with a sensor of this complexity. We validated the operation of the system first by simulating the Point-Spread Function of the HiRIS sensor, and compared these resulting PSFs to real-life measurements. Furthermore, we performed an ensonification experiment of a cluttered office environment, and generated B-mode images of the resulting datastreams.
\newline\newline
With HiRIS, we have developed a novel sensor system which is a step change in imaging capabilities of in-air sonar sensors, and which will allow virtually artifact-free imaging of real-world scenes. Therefore, we see the HiRIS as a virtual upper limit of in-air sonar imaging: more complex sensors could indeed be implemented, but the industrial relevance of systems of this complexity can be debated. Evidently, the approach we have taken during the development of HiRIS is in stark contrast to the developments of our eRTIS line of sensors~\cite{kerstens2019, verellen2020urtis}, during which component cost reduction was the major driving force during development. These sensors have been utilized in real-world applications under industrial constraints \cite{schenck2020airleakslam, schenck2019airleakslam, kerstens2023tracking, verellen2021beamforming, jansen2022real}, which has lead us to the ultimate question: what is the upper limit of ultrasound sensing that can be achieved, given the specular reflection model~\cite{pierce2019acoustics} under which the majority of ultrasound sensors operate. With HiRIS, we take the opposite approach: what is the upper limit that, given unrestricted sensing capabilities, can be achieved with in-air ultrasonic imaging, which in turn should lead to answers about the validity of the specular reflection model, the relative importance of diffraction echoes, and how semantic information about the environment is being translated into the ultrasonic sensing domain.
\newline\newline
To conclude, we believe that the HiRIS sensor will allow us to uncover the underlying mechanics of in-air ultrasound sensing in a previously unobtainable level of detail, which will then inform the development of future installments of 3D ultrasound sensors for industrial applications. 
\newline\newline
In future work, we aim to further quantify the performance of the HiRIS sensor, both in laboratory settings as well as real-world measurements. We will produce high-resolution datasets, which will be made open-source for the sensing community to evaluate and use. Using these datasets it should become possible to quantify how information-rich real-world ultrasound measurements really are, and how this information can be leveraged to provide robots with a rich understanding of their environments using ultrasound as a primary sensing modality. From these measurements, the effect of applying reduced-aperture microphone arrays instead of the large 1024 element array can be accurately calculated, as virtually any reduced aperture can be adequately simulated using the HiRIS array. 

\footnotesize
\section*{Acknowledgements}
The authors would like to thank the Bijzonder OnderzoeksFonds (BOF) of the University of Antwerp for funding this research project. 

\normalsize
\bibliography{citationLib}


\end{document}